\documentclass[prd,showpacs,preprintnumbers,amsmath,amssymb,superscriptaddress,nofootinbib]{revtex4}

\usepackage{graphicx}
\usepackage{dcolumn}
\usepackage{bm}
\usepackage{color}
\usepackage{subcaption}
\usepackage{pgfplots}
\pgfplotsset{compat=1.18}

\def\be{\begin{equation}}
\def\ee{\end{equation}}
\def\bea{\begin{eqnarray}}
\def\eea{\end{eqnarray}}

\allowdisplaybreaks[2]

\begin{document}

\title{Direct renormalization of flavor-changing currents in vNRQCD for $B_c^{(*)}$ decays}

\author{Jichen Pan\footnote{panjichen@pku.edu.cn}}
\affiliation{School of Physics, Peking University, Beijing 100871, China}
\affiliation{Center for High Energy Physics, Peking University, Beijing 100871, China}

\begin{abstract}
We present a direct vNRQCD calculation of the two-loop anomalous dimensions of the leading flavor-changing nonrelativistic currents that govern the leptonic decays of the $B_c$ and $B_c^*$ mesons. The calculation is performed at the matching scale $\nu=1$, where the anomalous dimensions are the ultraviolet counterterms required to cancel the residual infrared poles of the full-QCD short-distance coefficients. In contrast to the conventional extraction from those infrared poles, we determine the current renormalization constants directly in the effective theory. The pseudoscalar and vector channels are obtained from one spin-unified formula, with the spin dependence entering through $S(S+1)$. We also clarify the role of ultrasoft-induced mixing operators in the unequal-mass theory: they are essential for the full velocity running below $\nu=1$, but their Wilson coefficients vanish at the matching point and do not alter the fixed-order current anomalous dimensions reported here. The results agree with the known matching calculations and reduce smoothly to the equal-mass quarkonium limits.
\end{abstract}
\maketitle

\section{Introduction}

The $B_c$ meson stands as the unique ground-state meson within the Standard Model that is composed of two heavy quarks of distinct flavors. In this work, we adopt the convention $B_c^+(\bar b c)$, with the charge-conjugate channel implied throughout. In contrast to flavor-neutral quarkonia, such as charmonium and bottomonium, the $B_c$ possesses open bottom and charm quantum numbers, and thus decays exclusively via weak interactions. This distinctive property endows the $B_c$ with a relatively long lifetime, $\tau_{B_c}\approx0.51~{\rm ps}$ \cite{Workman:2022ynf,Navas:2024}, making it a valuable system for precision studies of QCD and weak interactions.

The $B_c$ meson was first observed by the CDF Collaboration in 1998 in $p\bar p$ collisions at $\sqrt{s}=1.8~{\rm TeV}$ via the semileptonic decay channel $B_c^\pm\to J/\psi\,\ell^\pm\nu$ \cite{CDF:1998ihx}. Subsequent measurements at the Tevatron and the LHC have refined its mass, lifetime, and production properties \cite{Workman:2022ynf,Navas:2024}. The spectroscopy of excited $B_c$ states has also witnessed substantial progress: the ATLAS Collaboration reported evidence for the $B_c(2S)$ state in 2014 \cite{ATLAS:2014lga}, and both CMS and LHCb later observed the $B_c(2S)$ and $B_c^*(2S)$ states \cite{CMS:2019uhm,LHCb:2019bem}. The ground-state vector meson $B_c^*(1S)$, however, has not yet been observed, primarily because it is expected to decay predominantly through the radiative transition $B_c^*\to B_c\gamma$, with a soft photon that poses significant experimental reconstruction challenges. The hyperfine splitting between the $B_c$ and $B_c^*$ is predicted to lie in the range of about $60$--$70~{\rm MeV}$ \cite{Eichten:1994,MartinGonzalez:2022}.

The double-heavy structure of the $B_c$ system permits a clean separation between perturbative dynamics at the heavy-quark mass scale and nonperturbative bound-state effects. In the NRQCD factorization approach \cite{Bodwin:1994jh}, the decay constants can be expressed as products of perturbatively calculable short-distance coefficients (SDCs) and universal long-distance matrix elements (LDMEs). The pseudoscalar decay constant $f_{B_c}$ governs the purely leptonic decay $B_c\to\ell\nu_\ell$ and enters the determination of the CKM matrix element $|V_{cb}|$ \cite{Braaten:1995cj,Chen:2015hva}. It is also relevant to precision flavor tests and searches for new physics beyond the Standard Model \cite{Patnaik:2024}. The vector decay constant $f_{B_c^*}$ plays the corresponding role for the vector meson $B_c^*$.

Considerable effort has been devoted to the higher-order QCD corrections to these decay constants. The one-loop QCD correction to $f_{B_c}$ was obtained by Braaten and Fleming \cite{Braaten:1995cj}. An important step toward the full higher-order result was taken by Onishchenko and Veretin, who computed the two-loop short-distance coefficient as an asymptotic expansion in the small mass ratio $m_c/m_b$ \cite{Onishchenko:2003ui}. Later, Chen and Qiao derived the complete two-loop result for an arbitrary mass ratio \cite{Chen:2015hva}, and the pseudoscalar coefficient is now known numerically through three loops \cite{Feng:2022yuf}. For the vector channel, the NNLO matching coefficient was obtained in Ref.~\cite{Tao:2022ttt}, and the $B_c^*$ decay constant was subsequently evaluated through ${\cal O}(\alpha_s^3)$ \cite{Sang:2022}. Three-loop matching coefficients are also available for a broad set of heavy flavor-changing currents \cite{Tao:2023a,Tao:2023b}. These studies indicate that large corrections to the matching coefficients can be partially compensated by corrections to the wave functions at the origin when the scale relations of potential NRQCD are employed \cite{Tao:2023a}. Complementary determinations from QCD sum rules and lattice QCD can be found in Refs.~\cite{Wang:2024,Colquhoun:2015}.

The mode separation adopted in this work is based on the threshold expansion, also known as the method of regions \cite{Beneke:1997zp}. Near a heavy-particle threshold, a loop momentum in QCD receives homogeneous contributions from the hard, soft, potential, and ultrasoft regions. Taylor expanding the integrand in each region yields the velocity expansion, while dimensional regularization avoids double counting via scaleless overlap integrals. Upon integrating out the hard region, one obtains NRQCD; retaining separate dynamical degrees of freedom for the remaining regions gives rise to velocity NRQCD (vNRQCD) \cite{Luke:2000,Manohar:2000a,Manohar:2000b,Hoang:2003a,Hoang:2003b,Rothstein:2018}.

vNRQCD separates the soft three-momentum scale from the ultrasoft kinetic-energy scale and correlates their renormalization through the velocity renormalization group (VRG). Its label formalism gives a homogeneous velocity expansion in dimensional regularization, makes the pull-up mechanism for mixed soft-ultrasoft counterterms explicit, and provides the natural framework for evolving the subtraction velocity from $\nu=1$ to $\nu\sim v$ \cite{Luke:2000,Hoang:2002,Hoang:2003b}. The framework has passed nontrivial checks in threshold-current evolution, the running of the QCD potential through order $v^2$, and ultrasoft-induced operator mixing \cite{Luke:2000,Manohar:2000a,Manohar:2000b,Hoang:2003b,Hoang:2004,Rothstein:2018}.

For the equal-mass case, the two-loop scale dependence of the currents has long been established within the vNRQCD framework. The first rigorous derivation was carried out by Luke and collaborators \cite{Luke:2000}, who performed a complete analysis for the $J/\psi$ channel. In this work, we extend this line of analysis to the unequal-mass sector. Our calculation provides a direct two-loop determination of the anomalous dimension for the flavor-changing currents, and treats the pseudoscalar and vector channels on an equal footing through a spin-unified formalism.

Our calculation is concerned with the two leading $S$-wave flavor-changing currents, $\chi_b^\dagger\psi_c$ and $\chi_b^\dagger\boldsymbol{\sigma}\psi_c$, which correspond to the pseudoscalar and vector channels, respectively. Rather than extracting their anomalous dimensions from the infrared poles left over after matching full QCD onto NRQCD, we compute directly the ultraviolet poles of the effective-theory current matrix elements. The main advance lies in the direct and spin-unified vNRQCD derivation, rather than in the numerical discovery of previously unknown anomalous dimensions. The outcome is a compact analytical formula valid for an arbitrary mass ratio $x=m_c/m_b$. In the equal-mass limit $x\to1$, our result reduces to the known quarkonium expressions, thereby providing a solid consistency check of the calculation.

It is significant to separate two related but conceptually different issues. The anomalous dimensions computed in this work are fixed-order matching-scale quantities: they are the counterterms that cancel the residual infrared poles in the full-QCD short-distance coefficients. A complete velocity renormalization group (VRG) evolution below the scale $\nu=1$ would further require the running of the potentials, which receives contributions from soft-loop contractions as well as ultrasoft-induced mixing operators. In this work we identify the relevant mode structure and the unequal-mass mixed operator induced by ultrasoft renormalization, but we do not attempt a full solution of the unequal-mass potential velocity running. This separation allows us to keep the present analysis focused on the current renormalization needed for the $B_c^{(*)}$ decay constants, while at the same time clarifying how our calculation fits into the broader vNRQCD program.

The remainder of this paper is organized as follows. In Sec.~II we extract the relevant mode content of vNRQCD from the threshold expansion, and present the ultrasoft, potential, and soft sectors of the effective Lagrangian. Section~III is devoted to the definition of the bare currents and their renormalization constants; we then evaluate the ultraviolet poles from potential-loop contributions and derive the two-loop anomalous dimensions. The running of the current-induced LDME is illustrated in Sec.~IV, and we conclude with a brief summary of our findings in Sec.~V.

\section{Threshold modes and vNRQCD}

\subsection{Threshold expansion}

The mode content of vNRQCD is fixed by the method of regions rather than postulated independently. Let $p$ denote the magnitude of the relative three-momentum of the $c\bar b$ pair and let $E\sim \frac{p^2}{2m}$ denote its nonrelativistic energy. Near threshold, the homogeneous momentum regions are \cite{Beneke:1997zp}
\begin{align}
\text{hard:}\quad       & k^0,|\mathbf k|\sim m_c\ \text{or}\ m_b,
\nonumber\\
\text{soft:}\quad       & k^0,|\mathbf k|\sim p,
\nonumber\\
\text{potential:}\quad  & k^0\sim E,\qquad |\mathbf k|\sim p,
\nonumber\\
\text{ultrasoft:}\quad  & k^0,|\mathbf k|\sim E.
\label{eq:threshold_regions}
\end{align}
The familiar notation $p\sim mv$ and $E\sim mv^2$ is only a shorthand for these scalings. In the unequal-mass calculation below, the two kinetic energies and all subleading mass factors are kept separately.

The threshold expansion is obtained by Taylor expanding each QCD integrand according to the scaling rules in Eq.~\eqref{eq:threshold_regions}, and integrating each expanded term over the full loop-momentum domain. In dimensional regularization, overlap contributions are scaleless. The hard region is absorbed into the Wilson coefficients when QCD is matched onto the nonrelativistic effective theory. Potential modes describe the heavy constituents, soft modes transfer momenta of order $p$, and ultrasoft modes carry the kinetic energy $E$. Conventional NRQCD contains these low-energy regions but does not make their homogeneous scaling manifest in a mass-independent subtraction scheme. vNRQCD implements the mode separation directly at the level of the fields \cite{Luke:2000,Hoang:2002}.

\subsection{Label fields and effective Lagrangian}

In the $c\bar b$ center-of-mass frame, the potential fields carry opposite soft three-momentum labels:
\begin{align}
p_c^\mu&=(m_c,\mathbf p)+k_c^\mu,
&
p_{\bar b}^\mu&=(m_b,-\mathbf p)+k_{\bar b}^\mu,
\label{eq:momentum_decomposition}
\end{align}
where $|\mathbf p|\sim p$ and $k_c^\mu,k_{\bar b}^\mu\sim E$. We denote the corresponding fields by $\psi_{\mathbf p}(x)$ and $\chi_{-\mathbf p}(x)$. Their coordinate dependence contains only residual ultrasoft momenta,
\begin{equation}
i\partial^\mu\psi_{\mathbf p}(x)\sim E\,\psi_{\mathbf p}(x),
\qquad
i\partial^\mu\chi_{-\mathbf p}(x)\sim E\,\chi_{-\mathbf p}(x).
\label{eq:potential_residual_scaling}
\end{equation}

Soft gauge, light-quark, and ghost fields have both a soft label and residual coordinate dependence. For example,
\begin{equation}
A_s^\mu(x)=\sum_q e^{-iq\cdot x}A_q^\mu(x),
\qquad q^\mu\sim p,\qquad
i\partial^\mu A_q^\nu(x)\sim E\,A_q^\nu(x),
\label{eq:soft_field_decomposition}
\end{equation}
with analogous decompositions for $\varphi_q$ and $c_q$. The continuous coordinate therefore labels only the residual ultrasoft momentum; the order-$p$ soft momentum is discrete in the label notation. Ultrasoft fields $A_u^\mu(x)$ carry no soft label.

The vNRQCD Lagrangian is organized as
\begin{equation}
{\cal L}_{\rm vNRQCD}={\cal L}_u+{\cal L}_p+{\cal L}_s,
\label{eq:vnrqcd_lagrangian}
\end{equation}
where the three terms contain ultrasoft interactions, four-fermion potentials, and soft interactions, respectively. Through the order needed here, the ultrasoft sector is
\begin{align}
{\cal L}_u={}&
\sum_{\mathbf p}\psi_{\mathbf p}^\dagger
\left[
iD^0-\frac{(\mathbf p-i\mathbf D)^2}{2m_c}
+\frac{\mathbf p^4}{8m_c^3}
\right]\psi_{\mathbf p}
\nonumber\\
&+\sum_{\mathbf p}\chi_{-\mathbf p}^\dagger
\left[
iD^0-\frac{(-\mathbf p-i\mathbf D)^2}{2m_b}
+\frac{\mathbf p^4}{8m_b^3}
\right]\chi_{-\mathbf p}
-\frac14F_{u\,\mu\nu}^AF_u^{A\mu\nu}+\cdots .
\label{eq:ultrasoft_lagrangian}
\end{align}
Here $D^\mu$ contains the ultrasoft gauge field, and the antiquark transforms with $\bar T^A=-(T^A)^*$. Reparameterization invariance requires the label and residual derivatives to occur in the combinations displayed in Eq.~\eqref{eq:ultrasoft_lagrangian}.

The potential sector has the form
\begin{equation}
{\cal L}_p=-\sum_{\mathbf p,\mathbf p'}
V(\mathbf p,\mathbf p')\,
\psi_{\mathbf p'}^\dagger\psi_{\mathbf p}\,
\chi_{-\mathbf p'}^\dagger\chi_{-\mathbf p}.
\label{eq:potential_lagrangian}
\end{equation}
For a $c\bar b$ system with unequal masses, the potential in the color basis used below can be written as \cite{Manohar:2000b,Hoang:2003b}
\begin{eqnarray}
V(\mathbf p,\mathbf p') &=& \left(T^A\otimes \bar T^A\right)
\Bigg[
\frac{V_c^{(T)}}{\mathbf k^2}
+ \frac{V_k^{(T)}\pi^2}{2m_r|\mathbf k|}
+\frac{V_2^{(T)}}{m_r^2}- \frac{V_{2o}^{(T)}(m_b-m_c)^2}{8m_b^2m_c^2}
+ \frac{V_r^{(T)}(\mathbf p^2+\mathbf p'^2)}{2m_bm_c\,\mathbf k^2} \nonumber\\
&&
+\frac{V_s^{(T)}}{m_bm_c}\mathbf S^2
+V_\Lambda^{(T)}\Lambda(\mathbf p',\mathbf p)
+\frac{V_t^{(T)}}{m_bm_c}T(\mathbf k)
\Bigg],
\label{eq:V_BC}
\end{eqnarray}
where $\mathbf k=\mathbf p'-\mathbf p$, $m_r=m_bm_c/(m_b+m_c)$, and $\mathbf S=(\boldsymbol{\sigma}_1+\boldsymbol{\sigma}_2)/2$. Here $\boldsymbol{\sigma}_1$ acts on the $\bar b$ line and $\boldsymbol{\sigma}_2$ on the $c$ line. The spin-dependent structures are defined by

\begin{equation}
\Lambda(\mathbf{p}',\mathbf{p})=
i\frac{\boldsymbol{\sigma}_1\cdot(\mathbf{p}'\times \mathbf{p})}{2\mathbf{k}^2m_b}
\left(\frac{1}{2m_b}+\frac{1}{m_c}\right)
+i\frac{\boldsymbol{\sigma}_2\cdot(\mathbf{p}'\times \mathbf{p})}{2\mathbf{k}^2m_c}
\left(\frac{1}{2m_c}+\frac{1}{m_b}\right),
\label{eq:Lambda_definition}
\end{equation}

\begin{equation}
T(\mathbf{k})=\boldsymbol{\sigma}_1 \cdot \boldsymbol{\sigma}_2
-\frac{3(\mathbf{k}\cdot \boldsymbol{\sigma}_1)
(\mathbf{k}\cdot \boldsymbol{\sigma}_2)}{\mathbf{k}^2}.
\label{eq:T_definition}
\end{equation}

At the matching scale $\nu=1$, the tree-level coefficients are
\begin{eqnarray}
V_c^{(T)}=V_r^{(T)}=V_{2o}^{(T)}=4\pi\alpha_s(m_b), \qquad
V_s^{(T)}=-\frac{4\pi}{3}\alpha_s(m_b), \nonumber\\
V_\Lambda^{(T)}=4\pi\alpha_s(m_b), \qquad
V_t^{(T)}=-\frac{\pi}{3}\alpha_s(m_b).
\label{eq:tree_coeffs}
\end{eqnarray}
where $V_{2}^{(T)}=0$, while $V_{2o}^{(T)}$ multiplies the contact term that vanishes in the equal-mass limit. The structures $\Lambda$ and $T$ do not contribute to the $S$-wave anomalous dimensions at the order considered here.

The Wilson coefficients in Eq.~\eqref{eq:V_BC} are matching coefficients of the vNRQCD potential. Using their known hard-scale values is part of specifying the effective theory, in the same way that the Coulomb coefficient is fixed by matching QCD onto vNRQCD. In particular, the unequal-mass coefficient of the $1/|\mathbf k|$ potential used below is taken from the QCD matching result of Ref.~\cite{Peset:2015vvi}; the subsequent extraction of the current ultraviolet counterterm is nevertheless carried out entirely inside vNRQCD.

For completeness, the soft sector contains both the free label fields and their interactions with potential quarks. Its structure can be displayed as
\begin{equation}
{\cal L}_s={\cal L}_s^{\rm free}+{\cal L}_s^{\rm int},
\label{eq:soft_lagrangian_split}
\end{equation}
with
\begin{align}
{\cal L}_s^{\rm free}=\sum_q\bigg[
&-\frac14\left(q^\mu A_q^\nu-q^\nu A_q^\mu\right)^\dagger
\left(q_\mu A_{q\nu}-q_\nu A_{q\mu}\right)
\nonumber\\
&+\bar\varphi_q\not q\,\varphi_q+\bar c_q\,q^2c_q
\bigg]+\cdots ,
\label{eq:soft_free_lagrangian}
\end{align}
where gauge-fixing terms are implicit. A convenient gauge-covariant organization of the potential-soft interactions is \cite{Luke:2000,Manohar:2000a,Rothstein:2018}
\begin{align}
{\cal L}_s^{\rm int}=-g_s^2
\sum_{\mathbf p,\mathbf p',q,q'}\bigg\{&
\frac12\psi_{\mathbf p'}^\dagger[A_q^\mu,A_{q'}^\nu]
U_{\mu\nu}^{(\sigma)}\psi_{\mathbf p}
+\frac12\psi_{\mathbf p'}^\dagger\{A_q^\mu,A_{q'}^\nu\}
W_{\mu\nu}^{(\sigma)}\psi_{\mathbf p}
\nonumber\\
&+\psi_{\mathbf p'}^\dagger[\bar c_{q'},c_q]Y^{(\sigma)}
\psi_{\mathbf p}
+\psi_{\mathbf p'}^\dagger T^A Z_\mu^{(\sigma)}
\psi_{\mathbf p}\,
\bar\varphi_{q'}T^A\gamma^\mu\varphi_q
\nonumber\\
&+\left(\psi\to\chi,\ T^A\to\bar T^A\right)
\bigg\}.
\label{eq:soft_interaction_lagrangian}
\end{align}
Momentum-label conservation in Eq.~\eqref{eq:soft_interaction_lagrangian} is implicit. A single soft field would transfer an energy of order $p$ and drive a potential quark off shell; the leading interactions therefore contain two soft fields. The coefficients have a homogeneous velocity expansion, for example $U_{\mu\nu}^{(\sigma)}=\sum_n U_{\mu\nu}^{(n)}$, with $U_{00}^{(0)}=1/q^0$ and $Z_0^{(0)}=1/(\mathbf p'-\mathbf p)^2$. Their complete expressions are given in Refs.~\cite{Luke:2000,Manohar:2000a}. Although no explicit soft field appears in the matching-scale diagrams evaluated below, this sector is indispensable for the running of the potentials and for soft-ultrasoft operator mixing away from $\nu=1$.

The VRG correlates the soft and ultrasoft subtraction scales as
\begin{equation}
\mu_S=m_b\nu,\qquad \mu_U=m_b\nu^2,
\label{eq:scale_correlation}
\end{equation}
where $m_b$ is chosen as the reference hard scale and $\nu$ is the subtraction velocity. This convention does not identify the two heavy masses: $m_c$ and $m_b$ remain explicit in the propagators and potentials. Running from $\nu=1$ to $\nu\sim v$ resums logarithms associated with both soft momentum and ultrasoft energy scales \cite{Luke:2000,Hoang:2003b}. In this work we use the mode-separated theory to compute the ultraviolet poles that renormalize the leading currents at the matching point $\nu=1$.

\section{Two-loop anomalous dimensions}

\subsection{Bare currents and renormalization convention}

We denote the mass ratio by
\begin{equation}
x\equiv\frac{m_c}{m_b}.
\label{eq:x_def}
\end{equation}
The bare leading $S$-wave current is
\begin{equation}
J_{S,B}=\sum_{\mathbf p}
\chi_{-\mathbf p}^\dagger\Gamma_S\psi_{\mathbf p},
\qquad
\Gamma_{S=0}=1,\qquad
\Gamma_{S=1}=\bm{\varepsilon}\cdot\bm{\sigma}.
\label{eq:bare_current}
\end{equation}
We first define its renormalized counterpart and only then introduce the anomalous dimension:
\begin{equation}
J_{S,B}=\widetilde Z_S(\mu_\Lambda)\,
J_S(\mu_\Lambda),\qquad
\widetilde Z_S=1+\delta\widetilde Z_S.
\label{eq:current_renormalization}
\end{equation}
The counterterm $\delta\widetilde Z_S$ is minus the ultraviolet pole part of the amputated vNRQCD vertex with an insertion of $J_S$. Since the bare operator is scale independent,
\begin{equation}
\gamma_S(\mu_\Lambda)
\equiv
\frac{d\ln\widetilde Z_S}{d\ln\mu_\Lambda}
=\left(\frac{\alpha_s}{\pi}\right)^2
\gamma_S^{(2)}(x)+{\cal O}(\alpha_s^3).
\label{eq:gamma_def}
\end{equation}
Equations~\eqref{eq:bare_current}--\eqref{eq:gamma_def} fix the sign and scale convention used throughout the paper.~\footnote{The full-QCD matching calculations in Refs.~\cite{Feng:2022yuf,Sang:2022}
define their anomalous dimensions with respect to $\ln\mu_\Lambda^2$.
Consequently, $\gamma_{S,{\rm here}}^{(2)}
=2\gamma_{S,{\rm match}}^{(2)}$. This factor of two is a derivative convention; it is unrelated to spin or to the two heavy-particle lines.}

The decay constants are defined by
\begin{subequations}
\label{eq:decay_constants}
\begin{align}
\langle0|\bar b\gamma^\mu\gamma_5 c|B_c(P)\rangle
&=iP^\mu f_{B_c},\\
\langle0|\bar b\gamma^\mu c|B_c^*(P)\rangle
&=M_{B_c^*}\varepsilon^\mu f_{B_c^*}.
\end{align}
\end{subequations}
At leading order in $v$, NRQCD factorization gives
\begin{equation}
f_H=\sqrt{\frac{M_H}{2}}\,
{\cal C}(m_b,x,\mu_\Lambda)\,
\langle0|J_S(\mu_\Lambda)|H(\mathbf P)\rangle .
\label{eq:factorization}
\end{equation}
Writing the same QCD current as
${\cal C}_B J_{S,B}={\cal C}(\mu_\Lambda)J_S(\mu_\Lambda)$ gives
${\cal C}_B={\cal C}\widetilde Z_S^{-1}$. Scale independence of the bare
operator then implies
\begin{equation}
\frac{d\ln \langle0|J_S(\mu_\Lambda)|H(\mathbf P)\rangle}
{d\ln\mu_\Lambda}
=-\left(\frac{\alpha_s}{\pi}\right)^2\gamma_S^{(2)}(x)
+{\cal O}(\alpha_s^3),
\label{eq:running_LDME}
\end{equation}
so that the scale dependences of the SDC and LDME cancel in
Eq.~\eqref{eq:factorization}.

\subsection{Modes at the matching point}

All ultraviolet poles below are evaluated at $\nu=1$, where
$\mu_S=\mu_U=m_b$; for the matching-scale current counterterm we identify
$\mu_\Lambda=\mu_S$. The fixed-order current counterterm at this point must be
distinguished from the complete velocity evolution for $\nu<1$. The former
cancels the residual infrared pole of the full-QCD SDC; the latter also
requires running potentials and operator mixing.

The leading coupling of an ultrasoft temporal gluon to a potential heavy
quark can be removed by the field redefinition
\begin{equation}
\psi_{\mathbf p}(x)\to W(x)\psi_{\mathbf p}(x),\qquad
W(x)=P\exp\left[
ig\int_{-\infty}^{0}d\lambda\, A_u^0(x+\lambda v)
\right],
\label{eq:us_wilson_line}
\end{equation}
with $v^\mu=(1,\mathbf 0)$ \cite{Luke:2000}. The corresponding quark and
antiquark Wilson lines cancel in a color-singlet leading current. Spatial
ultrasoft couplings and multipole-suppressed interactions enter at higher
orders in $v$.

For $\nu<1$, ultrasoft renormalization of the soft operators in
Eq.~\eqref{eq:soft_interaction_lagrangian} induces mixing operators through
the pull-up mechanism \cite{Hoang:2003b}. In the unequal-mass theory these
include line-diagonal terms proportional to $1/m_c^2$ and $1/m_b^2$ and
mixed terms proportional to $1/(m_cm_b)$. Their pull-up counterterms are
proportional to
$[\alpha_s(\mu_U)-\alpha_s(\mu_S)]/\epsilon$, so their Wilson coefficients
vanish at $\nu=1$. The $O(v^2)$ potentials also receive genuine soft-loop
contributions from contractions such as
$U_{00}^{(2)}U_{00}^{(0)}$ and
$U_{0i}^{(1)}U_{0i}^{(1)}$, together with light-quark and ghost operators
\cite{Manohar:2000a,Manohar:2000b}. These effects are required for complete
velocity running, but not for the matching-point counterterm calculated
here.

\subsection{Reduction of the potential loops}

An earlier pedagogical account of the vNRQCD setup and this potential-loop
calculation appears in Chap.~6 of Ref.~\cite{Pan:2023thesis}. We present the
derivation here in a self-contained form, keeping the two heavy masses
explicit throughout.

We now keep the two heavy propagators separate before carrying out the
energy integral. For a loop momentum $q^\mu=(q^0,\mathbf q)$ they are
\begin{equation}
G_{\bar b}(q)=
\frac{i}{q^0-\mathbf q^2/(2m_b)+i0},
\qquad
G_c(E-q)=
\frac{i}{E-q^0-\mathbf q^2/(2m_c)+i0}.
\label{eq:separate_heavy_propagators}
\end{equation}
The two poles lie on opposite sides of the real axis. Closing the contour
gives
\begin{align}
\int\frac{dq^0}{2\pi}\,
G_{\bar b}(q)G_c(E-q)
&=
\frac{i}{D(\mathbf q)},
\nonumber\\
D(\mathbf q)
&\equiv
E-\frac{\mathbf q^2}{2m_b}
-\frac{\mathbf q^2}{2m_c}+i0 .
\label{eq:energy_contour}
\end{align}
Only after this step can the sum in $D(\mathbf q)$ be written with a reduced
mass. We do not make that replacement in subleading insertions:
$\mathbf q^4/(8m_b^3)$ and $\mathbf q^4/(8m_c^3)$ remain distinct.

After the energy integrations, every diagram is an integral over
$d=3-2\epsilon$ spatial dimensions. We use
\begin{equation}
\int_{\mathbf q}\equiv
\int\frac{d^d\mathbf q}{(2\pi)^d}.
\label{eq:spatial_measure}
\end{equation}
The logarithmic ultraviolet poles reduce to
\begin{align}
I_0&\equiv
\mu_S^{4\epsilon}
\int_{\mathbf q,\mathbf q'}
\frac{1}{\mathbf q^2\mathbf q'^2(\mathbf q-\mathbf q')^2}
=\frac{1}{64\pi^2}\frac{1}{\epsilon_{\rm UV}}+\cdots ,
\nonumber\\
I_1&\equiv
\mu_S^{2\epsilon}
\int_{\mathbf q}\frac{1}{(\mathbf q^2)^{3/2}}
=\frac{1}{4\pi^2}\frac{1}{\epsilon_{\rm UV}}+\cdots .
\label{eq:master_uv_integrals}
\end{align}
For example, the nested subintegral in $I_0$ is
\begin{equation}
\int_{\mathbf q'}
\frac{1}{\mathbf q'^2(\mathbf q-\mathbf q')^2}
=
\frac{\Gamma(2-d/2)\Gamma^2(d/2-1)}
{(4\pi)^{d/2}\Gamma(d-2)}
(\mathbf q^2)^{d/2-2}.
\label{eq:massless_bubble}
\end{equation}
At $d=3$ its leading term is $1/(8|\mathbf q|)$. Retaining the
$\epsilon$ dependence of the power, the remaining radial integral behaves
as $\int^\infty dq\,q^{-1-4\epsilon}$ and gives the first pole in
Eq.~\eqref{eq:master_uv_integrals}; the one-loop integral behaves as
$\int^\infty dq\,q^{-1-2\epsilon}$ and gives the second.
An off-shell energy $E<0$ or nonzero external momentum is retained as an
infrared regulator while extracting these poles. Setting every external
scale to zero at the outset would turn $I_0$ and $I_1$ into scaleless
integrals and hide the separate ultraviolet and infrared poles.

For the color-singlet current,
\begin{equation}
(T^A\otimes\bar T^A)|1\rangle=-C_F|1\rangle .
\label{eq:singlet_projection}
\end{equation}
We denote the singlet-projected coefficient of each kinematic potential by
${\cal V}_i^{[1]}$. At tree level,
\begin{equation}
{\cal V}_c^{[1]}={\cal V}_r^{[1]}
={\cal V}_{2o}^{[1]}=-4\pi C_F\alpha_s,
\qquad
{\cal V}_s^{[1]}=\frac{4\pi}{3}C_F\alpha_s .
\label{eq:singlet_tree_potentials}
\end{equation}
With all amplitudes normalized to the tree current, minimal subtraction fixes
the two-loop counterterm through
\begin{equation}
\delta\widetilde Z_S\big|_{\alpha_s^2}
=-\left.
\left({\cal A}_a+{\cal A}_b+{\cal A}_c+{\cal A}_d
+{\cal A}_e+{\cal A}_f+{\cal A}_g\right)
\right|_{\rm UV}.
\label{eq:counterterm_from_diagrams}
\end{equation}
The anomalous dimension is then obtained by differentiating this
counterterm according to Eq.~\eqref{eq:gamma_def}, including the
$\mu_S$ factors displayed explicitly in the loop integrals. In particular,
the products of two tree-level potentials carry $\mu_S^{4\epsilon}$,
whereas the single insertion of the order-$\alpha_s^2$ potential is
normalized with $\mu_S^{2\epsilon}$. These factors generate the relative
factor of two between the first two classes and the $1/|\mathbf k|$ class
when the pole counterterms are differentiated. In the equal-mass limit,
this is precisely the relative normalization between the two-potential
terms and the ${\cal V}_{|\mathbf k|}/2$ term in Eq.~(55) of
Ref.~\cite{Luke:2000}.

\begin{figure}[t]
\centering
\includegraphics[width=0.85\linewidth]{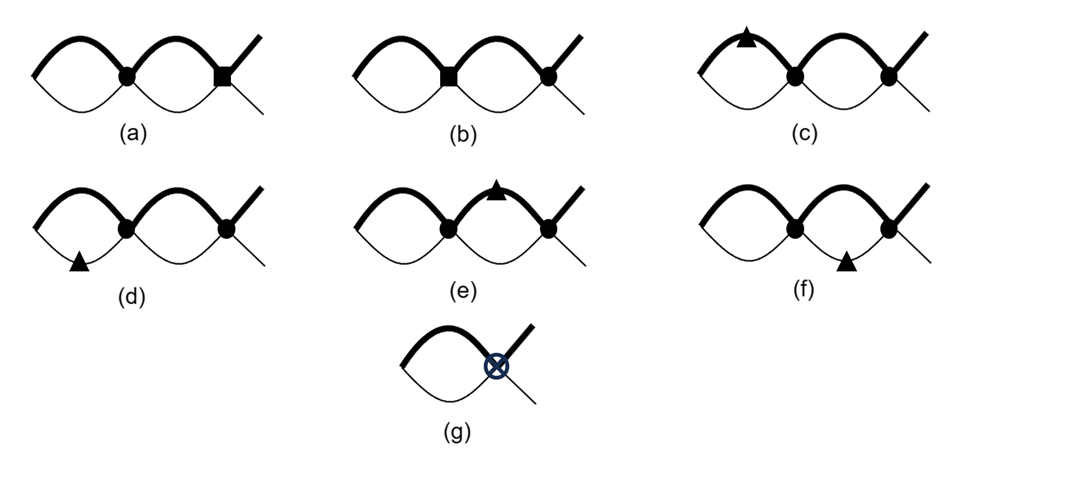}
\caption{Potential-loop topologies contributing to the two-loop current
anomalous dimensions. Solid dots denote Coulomb potentials, solid squares
denote the $V_r$, $V_{2o}$, and $V_s$ potentials, solid triangles denote
the separate $\mathbf p^4/(8m_b^3)$ or $\mathbf p^4/(8m_c^3)$ kinetic
insertions, and the crossed circle denotes the $1/|\mathbf k|$ potential.
Diagrams (a) and (b) give Eqs.~\eqref{eq:diagram_a_pole} and
\eqref{eq:diagram_b_pole}, diagrams (c)--(f) give
Eq.~\eqref{eq:kinetic_poles}, and diagram (g) gives
Eq.~\eqref{eq:diagram_g_pole}.}
\label{fig:potential_diagrams}
\end{figure}

Consider first diagrams (a) and (b) in
Fig.~\ref{fig:potential_diagrams}. After the two energy integrations,
diagram (a) is
\begin{align}
{\cal A}_a={}&
\mu_S^{4\epsilon}\int_{\mathbf q,\mathbf q'}
\frac{{\cal V}_c^{[1]}}
{D(\mathbf q)D(\mathbf q')(\mathbf q-\mathbf q')^2}
\bigg[
\frac{{\cal V}_r^{[1]}(\mathbf p^2+\mathbf q'^2)}
{2m_bm_c(\mathbf p-\mathbf q')^2}
\nonumber\\
&\hspace{35mm}
+\frac{{\cal V}_s^{[1]}S(S+1)}{m_bm_c}
-\frac{{\cal V}_{2o}^{[1]}(m_b-m_c)^2}
{8m_b^2m_c^2}
\bigg].
\label{eq:diagram_a_integral}
\end{align}
Expanding only the large-momentum part relevant to the logarithmic pole
gives
\begin{align}
{\cal A}_a\big|_{\rm UV}
=-\frac{4x}{(1+x)^2}{\cal V}_c^{[1]}
\bigg[
\frac{{\cal V}_r^{[1]}}{2}
+S(S+1){\cal V}_s^{[1]}
-\frac{(1-x)^2}{8x}{\cal V}_{2o}^{[1]}
\bigg]
\frac{1}{64\pi^2\epsilon_{\rm UV}},
\label{eq:diagram_a_pole}\\
{\cal A}_b\big|_{\rm UV}
=-\frac{4x}{(1+x)^2}{\cal V}_c^{[1]}
\frac{{\cal V}_r^{[1]}}{2}
\frac{1}{64\pi^2\epsilon_{\rm UV}}.
\label{eq:diagram_b_pole}
\end{align}
The factor $4x/(1+x)^2$ follows from the two unequal-mass denominators in
Eq.~\eqref{eq:energy_contour}; it is not inserted by a reduced-mass
replacement. Adding Eqs.~\eqref{eq:diagram_a_pole} and
\eqref{eq:diagram_b_pole} gives
\begin{align}
\left.({\cal A}_a+{\cal A}_b)\right|_{\rm UV}
&=-\frac{x}{16\pi^2(1+x)^2\epsilon_{\rm UV}}\,
{\cal V}_c^{[1]}
\left[
{\cal V}_r^{[1]}+S(S+1){\cal V}_s^{[1]}
-\frac{(1-x)^2}{8x}{\cal V}_{2o}^{[1]}
\right],
\nonumber\\
\delta\widetilde Z_{S,\rm pot}
&=\frac{x}{16\pi^2(1+x)^2\epsilon_{\rm UV}}\,
{\cal V}_c^{[1]}
\left[
{\cal V}_r^{[1]}+S(S+1){\cal V}_s^{[1]}
-\frac{(1-x)^2}{8x}{\cal V}_{2o}^{[1]}
\right].
\label{eq:potential_counterterm_steps}
\end{align}
For each tree-level potential insertion,
${\cal V}_{i,B}=\mu_S^{2\epsilon}{\cal V}_i$. Therefore, at fixed bare
potential coefficients,
\begin{equation}
\frac{d}{d\ln\mu_S}
\left({\cal V}_c^{[1]}{\cal V}_i^{[1]}\right)
=-4\epsilon\,
{\cal V}_c^{[1]}{\cal V}_i^{[1]}
+{\cal O}(\alpha_s^3).
\label{eq:two_potential_scale_derivative}
\end{equation}
At this order $\ln\widetilde Z_S=\delta\widetilde Z_S$, and hence
\begin{equation}
\gamma_{\rm pot}
=-4\epsilon\,\delta\widetilde Z_{S,\rm pot}
=-\frac{x}{4\pi^2(1+x)^2}\,
{\cal V}_c^{[1]}
\left[
{\cal V}_r^{[1]}+S(S+1){\cal V}_s^{[1]}
-\frac{(1-x)^2}{8x}{\cal V}_{2o}^{[1]}
\right].
\label{eq:gamma_potential_intermediate}
\end{equation}
Only now do we substitute the matching values in
Eq.~\eqref{eq:singlet_tree_potentials}. Comparing the result with
Eq.~\eqref{eq:gamma_def} yields
\begin{equation}
\gamma_{\rm pot}^{(2)}
=-\frac{4x\pi^2 C_F^2}{(1+x)^2}
\left[
1-\frac{S(S+1)}{3}-\frac{(1-x)^2}{8x}
\right].
\label{eq:gamma_potential_piece}
\end{equation}

Diagrams (c)--(f) contain one relativistic kinetic insertion. For example,
the energy integration in diagram (e), including the double pole generated
by the insertion on the anti-bottom line, gives
\begin{equation}
{\cal A}_e=
\mu_S^{4\epsilon}\int_{\mathbf q,\mathbf q'}
\frac{({\cal V}_c^{[1]})^2}{8m_b^3}
\frac{1}
{D(\mathbf q)[D(\mathbf q')]^2
(\mathbf q-\mathbf q')^2(\mathbf p-\mathbf q')^2}.
\label{eq:diagram_e_integral}
\end{equation}
Diagram (f) follows from $m_b\leftrightarrow m_c$. Their logarithmic pole
parts are
\begin{align}
{\cal A}_e\big|_{\rm UV}
&=-\frac{({\cal V}_c^{[1]})^2}{8m_b^3}
\left(\frac{2m_bm_c}{m_b+m_c}\right)^3
\frac{1}{64\pi^2\epsilon_{\rm UV}},
\nonumber\\
{\cal A}_f\big|_{\rm UV}
&=-\frac{({\cal V}_c^{[1]})^2}{8m_c^3}
\left(\frac{2m_bm_c}{m_b+m_c}\right)^3
\frac{1}{64\pi^2\epsilon_{\rm UV}},
\label{eq:kinetic_poles}
\end{align}
whereas ${\cal A}_c|_{\rm UV}={\cal A}_d|_{\rm UV}=0$. The latter two
topologies contain no logarithmic ultraviolet pole after the spatial
integrals are symmetrized; their remaining power divergences vanish in
dimensional regularization. Summing the two poles in
Eq.~\eqref{eq:kinetic_poles}
\begin{align}
\left.({\cal A}_e+{\cal A}_f)\right|_{\rm UV}
&=-\frac{({\cal V}_c^{[1]})^2}{64\pi^2\epsilon_{\rm UV}}
\frac{1-x+x^2}{(1+x)^2}.
\label{eq:kinetic_sum_pole}
\end{align}
By the same two-potential argument as in
Eq.~\eqref{eq:two_potential_scale_derivative},
\begin{equation}
\gamma_{\rm kin}
=-4\epsilon\left[
\frac{({\cal V}_c^{[1]})^2}{64\pi^2\epsilon_{\rm UV}}
\frac{1-x+x^2}{(1+x)^2}
\right]
=-\frac{({\cal V}_c^{[1]})^2}{16\pi^2}
\frac{1-x+x^2}{(1+x)^2}
=-\alpha_s^2C_F^2\frac{1-x+x^2}{(1+x)^2}.
\label{eq:gamma_kinetic_intermediate}
\end{equation}
Here the matching value of ${\cal V}_c^{[1]}$ is used only in the last
equality. Comparison with Eq.~\eqref{eq:gamma_def} then gives
\begin{equation}
\gamma_{\rm kin}^{(2)}
=-\pi^2C_F^2\frac{1-x+x^2}{(1+x)^2}.
\label{eq:gamma_kinetic_piece}
\end{equation}

Finally, diagram (g) contains a single $1/|\mathbf k|$ potential:
\begin{equation}
{\cal A}_g=
-\mu_S^{2\epsilon}\int_{\mathbf q}
\frac{1}{D(\mathbf q)}
\frac{\pi^2{\cal V}_k^{[1]}}
{2m_r|\mathbf q-\mathbf p|}.
\label{eq:diagram_g_integral}
\end{equation}
Using $D(\mathbf q)\to-\mathbf q^2/(2m_b)-\mathbf q^2/(2m_c)$ in the
ultraviolet region and Eq.~\eqref{eq:master_uv_integrals} gives
\begin{equation}
{\cal A}_g\big|_{\rm UV}
=\frac{{\cal V}_k^{[1]}}{4\epsilon_{\rm UV}}.
\label{eq:diagram_g_pole}
\end{equation}
We first keep ${\cal V}_k^{[1]}$ as an independent EFT Wilson coefficient.
There is only one potential insertion, so
${\cal V}_{k,B}^{[1]}=\mu_S^{2\epsilon}{\cal V}_k^{[1]}$ and
$d{\cal V}_k^{[1]}/d\ln\mu_S=-2\epsilon{\cal V}_k^{[1]}$ at fixed bare
coefficient. Equations~\eqref{eq:diagram_g_pole} and
\eqref{eq:counterterm_from_diagrams} therefore give
\begin{equation}
\delta\widetilde Z_{S,1/|\mathbf k|}
=-\frac{{\cal V}_k^{[1]}}{4\epsilon_{\rm UV}},
\qquad
\gamma_{1/|\mathbf k|}
=-2\epsilon\,\delta\widetilde Z_{S,1/|\mathbf k|}
=\frac{{\cal V}_k^{[1]}}{2}.
\label{eq:gamma_Vk_intermediate}
\end{equation}
Only after this differentiation do we insert the unequal-mass matching
coefficient
\begin{equation}
{\cal V}_k^{[1]}
=\alpha_s^2\left[
\frac{2x}{(1+x)^2}C_F^2-C_AC_F
\right],
\label{eq:Vk_singlet}
\end{equation}
taken from Ref.~\cite{Peset:2015vvi}. This coefficient is supplied when the
effective theory is defined; it is not extracted from the current
counterterm or from the full-QCD SDC. Substituting
Eq.~\eqref{eq:Vk_singlet} and comparing with Eq.~\eqref{eq:gamma_def} gives
\begin{equation}
\gamma_{1/|\mathbf k|}^{(2)}
=\pi^2\left[
\frac{x}{(1+x)^2}C_F^2-\frac{C_AC_F}{2}
\right].
\label{eq:gamma_Vk_piece}
\end{equation}

Adding these contributions yields a unified expression for the pseudoscalar and vector currents:~\footnote{In the equal-mass limit $x=1$, the $C_F^2$ term in Eq.~\eqref{eq:gamma_S_unified} reduces to the known quarkonium result. In that limit, the contribution from relativistic correction diagrams and the $C_F^2$ part of the graph $g$ cancel each other, so that a simplified treatment omitting the former may still reproduce the correct answer \cite{Chung:2020zqc}. For the unequal-mass case relevant to the $B_c$ system, however, this cancellation no longer holds, and a complete evaluation of all such contributions is required.}

\begin{equation}
\gamma_S^{(2)}(x)
=-\pi^2\left[
\frac{C_FC_A}{2}
+\frac{x^2+6x+1-\frac{8x}{3}S(S+1)}
{2(1+x)^2}C_F^2
\right].
\label{eq:gamma_S_unified}
\end{equation}

The spin dependence enters only through $\mathbf S^2=S(S+1)$, with $S=0$ for the pseudoscalar current and $S=1$ for the vector current. Equation~\eqref{eq:gamma_S_unified} is invariant under interchange of the two heavy masses, $x\to1/x$, as expected.

For the two channels of phenomenological interest, Eq.~\eqref{eq:gamma_S_unified} gives
\begin{eqnarray}
\gamma_{S=0}^{(2)}(x)
&=&-\pi^2\left[
\frac{C_FC_A}{2}
+\frac{1+6x+x^2}{2(1+x)^2}C_F^2
\right],
\label{eq:gamma_Bc}\\
\gamma_{S=1}^{(2)}(x)
&=&-\pi^2\left[
\frac{C_FC_A}{2}
+\frac{3x^2+2x+3}{6(1+x)^2}C_F^2
\right],
\label{eq:gamma_Bcstar}
\end{eqnarray}
corresponding to the $B_c$ and $B_c^*$ currents, respectively.

These results agree with the residual infrared poles obtained in full-QCD matching calculations. Equation~\eqref{eq:gamma_Bc} reproduces the two-loop anomalous dimension of the pseudoscalar current inferred from the $B_c$ leptonic decay amplitude \cite{Chen:2015hva,Feng:2022yuf}, while Eq.~\eqref{eq:gamma_Bcstar} reproduces the corresponding vector-current result \cite{Tao:2022ttt,Sang:2022}. To our knowledge, Eqs.~\eqref{eq:gamma_Bc} and \eqref{eq:gamma_Bcstar} give the first simultaneous direct vNRQCD derivation of the two unequal-mass matching-scale current anomalous dimensions in a spin-unified form \cite{Pan:2023thesis}. This narrower statement is compatible with the earlier equal-mass effective-theory calculations cited above.

As a further check, we take the equal-mass limit $x\to1$. Equation~\eqref{eq:gamma_S_unified} becomes
\begin{eqnarray}
\gamma_{S=0}^{(2)}(1)
&=&-\pi^2C_F\left(C_F+\frac{C_A}{2}\right),
\label{eq:equal_mass_ps}\\
\gamma_{S=1}^{(2)}(1)
&=&-\pi^2C_F\left(\frac{C_F}{3}+\frac{C_A}{2}\right),
\label{eq:equal_mass_vec}
\end{eqnarray}
in agreement with direct equal-mass effective-theory analyses and with the known matching results in the $d/d\ln\mu_\Lambda$ convention \cite{Luke:2000,Chung:2020zqc,Chen:2015hva,Tao:2022ttt}.\footnote{In the $d/d\ln\mu_\Lambda^2$ convention of the matching papers, the two expressions are divided by two and become $-\pi^2C_F(C_F/2+C_A/4)$ and $-\pi^2C_F(C_F/6+C_A/4)$, respectively.} This smooth limit provides a stringent check of both the unequal-mass formula and the convention conversion.

We finally stress the scope of the result. The calculation is performed at the matching scale $\nu=1$. Below this scale, ultrasoft renormalization of soft vertices induces additional operators, including unequal-mass mixed structures and the sum operators ${\cal O}_{k1}^{(S)}$ and ${\cal O}_{k2}^{(S)}$ in the notation of Ref.~\cite{Hoang:2003b}. Their Wilson coefficients vanish at $\nu=1$ but are generated by VRG evolution. A complete treatment of the potential running for $\nu<1$ also requires the soft-loop contractions discussed above. This full running problem is beyond the scope of the present work. The anomalous dimensions in Eqs.~\eqref{eq:gamma_Bc} and \eqref{eq:gamma_Bcstar} are the fixed-order matching-scale current anomalous dimensions and are precisely the quantities needed to cancel the residual infrared poles in the full-QCD SDCs.

\section{Illustrative running of the long-distance matrix elements}

We now turn to the numerical scale dependence induced by the current anomalous dimension in Eq.~\eqref{eq:gamma_S_unified}. This exercise accounts for the evolution of the leading current only, and does not incorporate the running of the potentials or the ultrasoft-induced mixing of operators, both of which would be needed for a complete VRG evolution below $\nu=1$.

The HPQCD decay constants are continuum-limit QCD observables extracted with a specific lattice-current normalization and extrapolation procedure \cite{McNeile:2012qf,Colquhoun:2015}. As such, they carry no intrinsic $\overline{\rm MS}$ NRQCD factorization scale. Converting an absolute lattice matrix element into the vNRQCD LDME in Eq.~\eqref{eq:factorization} would require a scheme-matched short-distance coefficient. To avoid assigning an unjustified scale to the lattice normalization, we work instead with the dimensionless ratio
\begin{equation}
R_H(\mu,\mu_1)
\equiv
\frac{\langle0|J_S(\mu)|H(\mathbf P)\rangle}
{\langle0|J_S(\mu_1)|H(\mathbf P)\rangle}.
\label{eq:ldme_ratio_def}
\end{equation}
This normalization removes the absolute lattice-current matching ambiguity
and isolates the running induced by the anomalous dimension at the stated
perturbative accuracy. Residual scheme and scale dependence beyond this
accuracy remains part of the perturbative truncation uncertainty.

We use the standard potential-model-scale masses
\begin{equation}
m_c=1.50~{\rm GeV},\qquad m_b=4.80~{\rm GeV},
\label{eq:running_masses}
\end{equation}
which are representative of heavy-quark phenomenology \cite{Eichten:1994}, and
choose
\begin{equation}
\mu_1=\sqrt{m_bm_c}=2.683~{\rm GeV},
\qquad
\mu_2=m_c=1.50~{\rm GeV}.
\label{eq:running_scales}
\end{equation}
The geometric mean treats the two heavy masses symmetrically and sets to zero
the fixed-order factorization logarithm
$\ln[\mu_\Lambda^2/(m_bm_c)]$ appearing in the unequal-mass short-distance
coefficient \cite{Sang:2022}. It also provides a natural intermediate scale
between the bottom and charm masses, consistent with the HPQCD observation
that the physical $B_c$ system exhibits features of both heavy-heavy and
heavy-light dynamics \cite{McNeile:2012qf}. We take $\mu_2=m_c$ as the
terminal scale; the interval remains perturbative in both fixed-flavor
coupling conventions considered below.

At one-loop accuracy we use
\begin{equation}
\frac{1}{\alpha_s^{(n_f)}(\mu)}
=
\frac{1}{\alpha_s^{(n_f)}(\mu_0)}
+\frac{\beta_0^{(n_f)}}{2\pi}\ln\frac{\mu}{\mu_0},
\qquad
\beta_0^{(n_f)}=11-\frac{2n_f}{3}.
\label{eq:alpha_one_loop}
\end{equation}
To facilitate combination with SDCs quoted in either a three- or four-flavor
coupling convention, we perform the illustrative evolution for both $n_f=3$
and $n_f=4$, keeping the selected value fixed over the full interval
$\mu_2\leq\mu\leq\mu_1$. The two couplings are normalized consistently to the
same world-average high-scale input, $\alpha_s^{(5)}(m_Z)=0.118$
\cite{Navas:2024}.

Combining Eq.~\eqref{eq:running_LDME} with one-loop running of the coupling yields
\begin{equation}
R_H(\mu,\mu_1)
=
\exp\left[
+\frac{2\gamma_S^{(2)}(x)}{\beta_0^{(n_f)}\pi}
\left(\alpha_s^{(n_f)}(\mu)-\alpha_s^{(n_f)}(\mu_1)\right)
\right].
\label{eq:ldme_evolution}
\end{equation}
Equation~\eqref{eq:ldme_evolution} resums the leading logarithms generated by
the ${\cal O}(\alpha_s^2)$ current anomalous dimension with one-loop coupling
running; its fixed-order expansion begins at
${\cal O}[\alpha_s^2\ln(\mu/\mu_1)]$. Here $S$ denotes the spin of
$H=B_c$ or $B_c^*$. For $x=m_c/m_b=0.3125$, Eqs.~\eqref{eq:gamma_Bc} and
\eqref{eq:gamma_Bcstar} give
\begin{equation}
\gamma_{S=0}^{(2)}=-34.8781,
\qquad
\gamma_{S=1}^{(2)}=-26.3902.
\label{eq:running_gammas}
\end{equation}
The corresponding endpoint ratios are
\begin{subequations}
\label{eq:running_endpoint_ratios}
\begin{equation}
R_{B_c}(\mu_2,\mu_1)
=
\begin{cases}
0.8629, & n_f=3,\\
0.8609, & n_f=4,
\end{cases}
\label{eq:running_endpoint_ratio_Bc}
\end{equation}
\begin{equation}
R_{B_c^*}(\mu_2,\mu_1)
=
\begin{cases}
0.8944, & n_f=3,\\
0.8929, & n_f=4.
\end{cases}
\label{eq:running_endpoint_ratio_Bcstar}
\end{equation}
\end{subequations}
Thus, within this current-only leading-logarithmic evolution, lowering the scale from
$2.683$ to $1.50~{\rm GeV}$ decreases the pseudoscalar LDME by about
$13.7\%$ ($13.9\%$) and the vector LDME by about $10.6\%$ ($10.7\%$) for
$n_f=3$ ($n_f=4$). The larger pseudoscalar effect
reflects the greater magnitude of $\gamma_{S=0}^{(2)}$. These percentages
quantify the current anomalous-dimension contribution and should not be
interpreted as complete velocity-running corrections.

\begin{figure*}[t]
\centering
\begin{subfigure}[t]{0.49\textwidth}
\centering
\begin{tikzpicture}
\begin{axis}[
width=0.92\linewidth,
height=0.62\linewidth,
xmin=1.50,xmax=2.70,
ymin=0.84,ymax=1.01,
xlabel={$\mu~[{\rm GeV}]$},
ylabel={$R_{B_c}(\mu,\mu_1)$},
legend style={font=\scriptsize,draw=none,fill=none,at={(0.03,0.97)},anchor=north west},
tick label style={font=\scriptsize},
label style={font=\scriptsize},
grid=major,
grid style={dotted,gray!45}
]
\addplot[blue,very thick,domain=1.50:2.6832816,samples=160]
{exp(2*(-34.8780729)/(8.3333333333*pi)*
((0.2432714142/(1+0.2432714142*8.3333333333/(2*pi)*ln(x/2.683281573)))
-0.2432714142))};
\addlegendentry{$n_f=4$}
\addplot[orange!85!black,dashed,very thick,domain=1.50:2.6832816,samples=160]
{exp(2*(-34.8780729)/(9*pi)*
((0.2396473/(1+0.2396473*9/(2*pi)*ln(x/2.683281573)))
-0.2396473))};
\addlegendentry{$n_f=3$}
\addplot[black,dotted,thick] coordinates {(2.683281573,0.84) (2.683281573,1.01)};
\addlegendentry{$\mu_1=\sqrt{m_bm_c}$}
\addplot[blue,only marks,mark=*,mark size=2.2pt] coordinates {(2.683281573,1.0)};
\end{axis}
\end{tikzpicture}
\caption{Pseudoscalar $B_c$ channel.}
\label{fig:ldme_running_Bc}
\end{subfigure}
\hfill
\begin{subfigure}[t]{0.49\textwidth}
\centering
\begin{tikzpicture}
\begin{axis}[
width=0.92\linewidth,
height=0.62\linewidth,
xmin=1.50,xmax=2.70,
ymin=0.88,ymax=1.01,
xlabel={$\mu~[{\rm GeV}]$},
ylabel={$R_{B_c^*}(\mu,\mu_1)$},
legend style={font=\scriptsize,draw=none,fill=none,at={(0.03,0.97)},anchor=north west},
tick label style={font=\scriptsize},
label style={font=\scriptsize},
grid=major,
grid style={dotted,gray!45}
]
\addplot[red,very thick,domain=1.50:2.6832816,samples=160]
{exp(2*(-26.3902297)/(8.3333333333*pi)*
((0.2432714142/(1+0.2432714142*8.3333333333/(2*pi)*ln(x/2.683281573)))
-0.2432714142))};
\addlegendentry{$n_f=4$}
\addplot[teal!80!black,dashed,very thick,domain=1.50:2.6832816,samples=160]
{exp(2*(-26.3902297)/(9*pi)*
((0.2396473/(1+0.2396473*9/(2*pi)*ln(x/2.683281573)))
-0.2396473))};
\addlegendentry{$n_f=3$}
\addplot[black,dotted,thick] coordinates {(2.683281573,0.88) (2.683281573,1.01)};
\addlegendentry{$\mu_1=\sqrt{m_bm_c}$}
\addplot[red,only marks,mark=square*,mark size=2.2pt] coordinates {(2.683281573,1.0)};
\end{axis}
\end{tikzpicture}
\caption{Vector $B_c^*$ channel.}
\label{fig:ldme_running_Bcstar}
\end{subfigure}
\caption{Current-induced running ratios for the leading $B_c$ and $B_c^*$
LDMEs, generated by Eqs.~\eqref{eq:gamma_Bc} and
\eqref{eq:gamma_Bcstar}. The parameters are $m_c=1.50~{\rm GeV}$,
$m_b=4.80~{\rm GeV}$, and $\mu_1=2.683~{\rm GeV}$. The solid and dashed
curves use the $n_f=4$ and $n_f=3$ coupling conventions, respectively; the
dotted vertical line marks $\mu_1$.}
\label{fig:ldme_running}
\end{figure*}
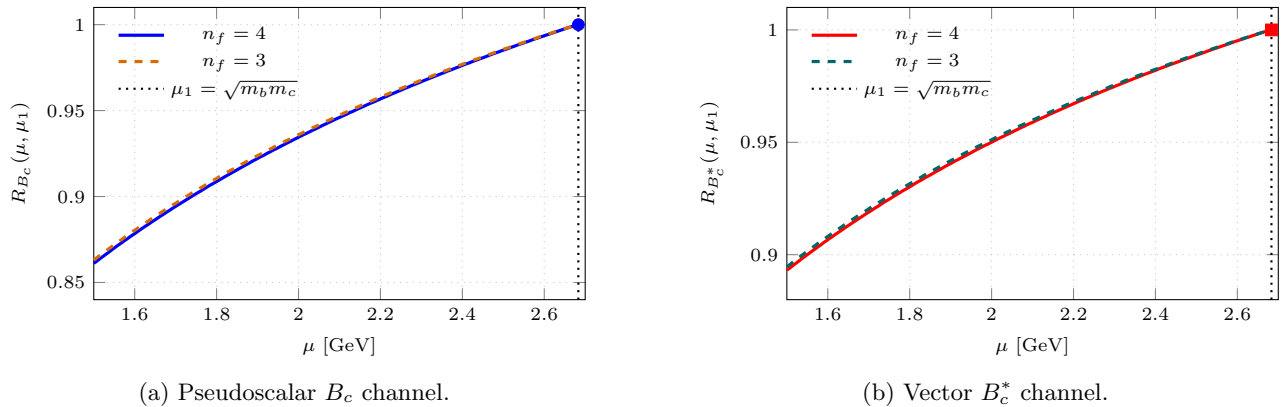

\section{Conclusion}

We have directly calculated the two-loop anomalous dimensions of the leading flavor-changing NRQCD currents relevant to the leptonic decays of the $B_c$ and $B_c^*$ mesons. The calculation was carried out in vNRQCD at the matching scale $\nu=1$, and yields a single unified expression in Eq.~\eqref{eq:gamma_S_unified} for the pseudoscalar and vector channels, which are distinguished by the spin factor $S(S+1)$.

Our results agree with the anomalous dimensions extracted from residual infrared poles in full-QCD matching calculations \cite{Chen:2015hva,Tao:2022ttt,Sang:2022}, and reduce smoothly to the known equal-mass effective-theory results \cite{Luke:2000,Chung:2020zqc}. These agreements provide a nontrivial check of NRQCD factorization for flavor-changing currents at two-loop order, and confirm that vNRQCD offers a self-contained description of the current renormalization at the matching point. The methodological advance of this work is a direct unequal-mass derivation that treats the pseudoscalar and vector channels simultaneously, and explicitly demonstrates how the mode separation of vNRQCD controls the calculation.

An immediate extension would be to push the analysis below $\nu=1$ by including the full unequal-mass velocity running of the potentials, the ultrasoft-induced operators and corresponding pull-up counterterms, the line-diagonal wavefunction completion, and the genuine soft-loop contractions of subleading soft operators. It would also be worthwhile to apply the same direct vNRQCD method to higher-order relativistic corrections, such as the $O(\alpha_s^2v^2)$ contributions to the flavor-changing currents.

\begin{acknowledgments}
I thank Prof. Yu Jia for suggesting this topic, for carefully reading the manuscript, and for providing valuable suggestions on its revision.
\end{acknowledgments}

\bibliography{Bcref}

@article{Navas:2024,
    author = {Navas, S. and others},
    collaboration = {Particle Data Group},
    title = {Review of Particle Physics},
    journal = {Phys. Rev. D},
    volume = {110},
    number = {3},
    pages = {030001},
    year = {2024},
    doi = {10.1103/PhysRevD.110.030001},
    reportNumber = {PDG-2024},
    SLACcitation = {%%CITATION = PHRVA,D110,030001;%%}
}

@article{CDF:1998ihx,
    author = {Abe, F. and others},
    collaboration = {CDF},
    title = {Observation of the $B_c$ Meson in $p\bar p$ Collisions at $\sqrt{s} = 1.8$ TeV},
    journal = {Phys. Rev. Lett.},
    volume = {81},
    pages = {2432--2437},
    year = {1998},
    doi = {10.1103/PhysRevLett.81.2432},
    eprint = {hep-ex/9805034},
    archivePrefix = {arXiv},
    primaryClass = {hep-ex},
    reportNumber = {FERMILAB-PUB-98/157-E}
}

@article{ATLAS:2014lga,
    author = {Aad, G. and others},
    collaboration = {ATLAS},
    title = {Observation of an Excited $B_c^\pm$ Meson State with the ATLAS Detector},
    journal = {Phys. Rev. Lett.},
    volume = {113},
    number = {21},
    pages = {212004},
    year = {2014},
    doi = {10.1103/PhysRevLett.113.212004},
    eprint = {1407.1032},
    archivePrefix = {arXiv},
    primaryClass = {hep-ex},
    reportNumber = {CERN-PH-EP-2014-137},
    SLACcitation = {%%CITATION = ARXIV:1407.1032;%%}
}

@article{CMS:2019uhm,
    author = {Sirunyan, A. M. and others},
    collaboration = {CMS},
    title = {Observation of Two Excited $B_c^+$ States and Measurement of the $B_c^+(2S)$ Mass in $pp$ Collisions at $\sqrt{s} = 13$ TeV},
    journal = {Phys. Rev. Lett.},
    volume = {122},
    number = {13},
    pages = {132001},
    year = {2019},
    doi = {10.1103/PhysRevLett.122.132001},
    eprint = {1902.00571},
    archivePrefix = {arXiv},
    primaryClass = {hep-ex},
    reportNumber = {CMS-BPH-18-007, CERN-EP-2019-014},
    SLACcitation = {%%CITATION = ARXIV:1902.00571;%%}
}

@article{LHCb:2019bem,
    author = {Aaij, R. and others},
    collaboration = {LHCb},
    title = {Observation of an Excited $B_c^+$ State},
    journal = {Phys. Rev. Lett.},
    volume = {122},
    number = {23},
    pages = {232001},
    year = {2019},
    doi = {10.1103/PhysRevLett.122.232001},
    eprint = {1904.00081},
    archivePrefix = {arXiv},
    primaryClass = {hep-ex},
    reportNumber = {LHCb-PAPER-2019-007, CERN-EP-2019-050},
    SLACcitation = {%%CITATION = ARXIV:1904.00081;%%}
}

@article{Eichten:1994,
    author = {Eichten, Estia J. and Quigg, Chris},
    title = {Mesons with beauty and charm: Spectroscopy},
    journal = {Phys. Rev. D},
    volume = {49},
    pages = {5845--5856},
    year = {1994},
    doi = {10.1103/PhysRevD.49.5845},
    eprint = {hep-ph/9402210},
    archivePrefix = {arXiv},
    primaryClass = {hep-ph},
    reportNumber = {FERMILAB-PUB-94-032-T},
    SLACcitation = {%%CITATION = HEP-PH/9402210;%%}
}

@article{MartinGonzalez:2022,
    author = {Mart\'{i}n-Gonz\'{a}lez, B. and Ortega, P. G. and Entem, D. R. and Fern\'{a}ndez, F. and Segovia, J.},
    title = {Toward the discovery of novel $B_c$ states: Radiative and hadronic transitions},
    journal = {Phys. Rev. D},
    volume = {106},
    number = {5},
    pages = {054009},
    year = {2022},
    doi = {10.1103/PhysRevD.106.054009},
    eprint = {2205.05950},
    archivePrefix = {arXiv},
    primaryClass = {hep-ph},
    SLACcitation = {%%CITATION = ARXIV:2205.05950;%%}
}

@article{Bodwin:1994jh,
    author = {Bodwin, Geoffrey T. and Braaten, Eric and Lepage, G. Peter},
    title = {Rigorous QCD analysis of inclusive annihilation and production of heavy quarkonium},
    journal = {Phys. Rev. D},
    volume = {51},
    pages = {1125--1171},
    year = {1995},
    doi = {10.1103/PhysRevD.51.1125},
    note = {[Erratum: Phys. Rev. D 55, 5853 (1997)]},
    eprint = {hep-ph/9407339},
    archivePrefix = {arXiv},
    primaryClass = {hep-ph},
    reportNumber = {ANL-HEP-PR-94-24},
    SLACcitation = {%%CITATION = HEP-PH/9407339;%%}
}

@article{Braaten:1995cj,
    author = {Braaten, Eric and Fleming, Sean},
    title = {QCD Radiative Corrections to the Leptonic Decay Rate of the $B_c$ Meson},
    journal = {Phys. Rev. D},
    volume = {52},
    pages = {181--185},
    year = {1995},
    doi = {10.1103/PhysRevD.52.181},
    eprint = {hep-ph/9501296},
    archivePrefix = {arXiv},
    primaryClass = {hep-ph},
    reportNumber = {NUHEP-TH-95-1},
    SLACcitation = {%%CITATION = HEP-PH/9501296;%%}
}

@article{Onishchenko:2003ui,
    author = {Onishchenko, Andrei I. and Veretin, Oleg L.},
    title = {Two-loop QCD corrections to the leptonic constant of the $B_c$-meson},
    journal = {Eur. Phys. J. C},
    volume = {50},
    pages = {801--808},
    year = {2007},
    doi = {10.1140/epjc/s10052-007-0255-1},
    eprint = {hep-ph/0302132},
    archivePrefix = {arXiv},
    primaryClass = {hep-ph},
    reportNumber = {WSU-HEP-0302, TTP-03-05},
    SLACcitation = {%%CITATION = HEP-PH/0302132;%%}
}

@article{Chen:2015hva,
    author = {Chen, Long-Bin and Qiao, Cong-Feng},
    title = {Two-loop QCD corrections to $B_c$ meson leptonic decays},
    journal = {Phys. Lett. B},
    volume = {748},
    pages = {443--450},
    year = {2015},
    doi = {10.1016/j.physletb.2015.07.043},
    eprint = {1503.05122},
    archivePrefix = {arXiv},
    primaryClass = {hep-ph},
    SLACcitation = {%%CITATION = ARXIV:1503.05122;%%}
}

@article{Feng:2022yuf,
    author = {Feng, Feng and Jia, Yu and Mo, Zhewen and Pan, Jichen and Sang, Wen-Long and Zhang, Jia-Yue},
    title = {Three-loop QCD corrections to the decay constant of $B_c$},
    year = {2022},
    eprint = {2208.04302},
    archivePrefix = {arXiv},
    primaryClass = {hep-ph},
    SLACcitation = {%%CITATION = ARXIV:2208.04302;%%}
}

@article{Tao:2022ttt,
    author = {Tao, Wei and Zhu, Ruilin and Xiao, Zhen-Jun},
    title = {Next-to-next-to-leading order matching of beauty-charmed meson $B_c$ and $B_c^*$ decay constants},
    journal = {Phys. Rev. D},
    volume = {106},
    number = {11},
    pages = {114037},
    year = {2022},
    doi = {10.1103/PhysRevD.106.114037},
    eprint = {2209.15521},
    archivePrefix = {arXiv},
    primaryClass = {hep-ph},
    SLACcitation = {%%CITATION = ARXIV:2209.15521;%%}
}

@article{Sang:2022,
    author = {Sang, Wen-Long and Zhang, Hong-Fei and Zhou, Ming-Zhen},
    title = {Decay constant of $B_c^*$ accurate up to $\mathcal{O}(\alpha_s^3)$},
    journal = {Phys. Lett. B},
    volume = {839},
    pages = {137812},
    year = {2023},
    doi = {10.1016/j.physletb.2023.137812},
    eprint = {2210.02979},
    archivePrefix = {arXiv},
    primaryClass = {hep-ph},
    SLACcitation = {%%CITATION = ARXIV:2210.02979;%%}
}

@article{Tao:2023a,
    author = {Tao, Wei and Xiao, Zhen-Jun and Zhu, Ruilin},
    title = {Three-loop matching coefficients for heavy flavor-changing currents and the phenomenological applications},
    journal = {JHEP},
    volume = {05},
    pages = {189},
    year = {2023},
    doi = {10.1007/JHEP05(2023)189},
    eprint = {2303.07220},
    archivePrefix = {arXiv},
    primaryClass = {hep-ph},
    SLACcitation = {%%CITATION = ARXIV:2303.07220;%%}
}

@article{Tao:2023b,
    author = {Tao, Wei and Xiao, Zhen-Jun},
    title = {Decay constants of $c\bar b$ mesons involving the ten heavy flavor-changing currents at N$^3$LO QCD},
    journal = {JHEP},
    volume = {06},
    pages = {012},
    year = {2024},
    doi = {10.1007/JHEP06(2024)012},
    eprint = {2310.17500},
    archivePrefix = {arXiv},
    primaryClass = {hep-ph},
    SLACcitation = {%%CITATION = ARXIV:2310.17500;%%}
}

@article{Wang:2024,
    author = {Wang, Zhi-Gang},
    title = {The $B_c$ meson and its scalar cousin with the QCD sum rules},
    journal = {Chin. Phys. C},
    volume = {48},
    pages = {103104},
    year = {2024},
    doi = {10.1088/1674-1137/ad5a71},
    eprint = {2401.12571},
    archivePrefix = {arXiv},
    primaryClass = {hep-ph},
    SLACcitation = {%%CITATION = ARXIV:2401.12571;%%}
}

@article{Chung:2020zqc,
    author = {Chung, Hee Sok},
    title = {$\overline{\rm MS}$ renormalization of $S$-wave quarkonium wavefunctions at the origin},
    journal = {JHEP},
    volume = {12},
    pages = {065},
    year = {2020},
    doi = {10.1007/JHEP12(2020)065},
    eprint = {2007.01737},
    archivePrefix = {arXiv},
    primaryClass = {hep-ph},
    reportNumber = {TUM-EFT-135-20},
    SLACcitation = {%%CITATION = ARXIV:2007.01737;%%}
}

@article{Patnaik:2024,
    author = {Patnaik, Sonali},
    title = {Unravelling theoretical challenges in understanding $B_c$ meson decay},
    eprint = {2411.11413},
    archivePrefix = {arXiv},
    primaryClass = {hep-ph},
    year = {2024},
    SLACcitation = {%%CITATION = ARXIV:2411.11413;%%}
}

@article{Colquhoun:2015,
    author = {Colquhoun, B. and Davies, C. T. H. and Dowdall, R. J. and Kettle, J. and Koponen, J. and Lepage, G. P. and Lytle, A. T.},
    collaboration = {HPQCD},
    title = {$B$-meson decay constants: a more complete picture from full lattice QCD},
    journal = {Phys. Rev. D},
    volume = {91},
    number = {11},
    pages = {114509},
    year = {2015},
    doi = {10.1103/PhysRevD.91.114509},
    eprint = {1503.05762},
    archivePrefix = {arXiv},
    primaryClass = {hep-lat},
    SLACcitation = {%%CITATION = ARXIV:1503.05762;%%}
}

@article{McNeile:2012qf,
    author = {McNeile, C. and Davies, C. T. H. and Follana, E. and Hornbostel, K. and Lepage, G. P.},
    collaboration = {HPQCD},
    title = {Heavy meson masses and decay constants from relativistic heavy quarks in full lattice QCD},
    journal = {Phys. Rev. D},
    volume = {86},
    pages = {074503},
    year = {2012},
    doi = {10.1103/PhysRevD.86.074503},
    eprint = {1207.0994},
    archivePrefix = {arXiv},
    primaryClass = {hep-lat},
    SLACcitation = {%%CITATION = ARXIV:1207.0994;%%}
}

@article{Workman:2022ynf,
    author = {Workman, R. L. and others},
    collaboration = {Particle Data Group},
    title = {Review of Particle Physics},
    journal = {PTEP},
    volume = {2022},
    pages = {083C01},
    year = {2022},
    doi = {10.1093/ptep/ptac097},
    reportNumber = {PDG-2022},
    SLACcitation = {%%CITATION = PTEP,2022,083C01;%%}
}

@article{Beneke:1997zp,
    author = {Beneke, Martin and Smirnov, Vladimir A.},
    title = {Asymptotic expansion of Feynman integrals near threshold},
    journal = {Nucl. Phys. B},
    volume = {522},
    pages = {321--344},
    year = {1998},
    doi = {10.1016/S0550-3213(98)00138-2},
    eprint = {hep-ph/9711391},
    archivePrefix = {arXiv},
    primaryClass = {hep-ph},
    reportNumber = {CERN-TH-97-315},
    SLACcitation = {%%CITATION = HEP-PH/9711391;%%}
}

@article{Luke:2000,
    author = {Luke, Michael E. and Manohar, Aneesh V. and Rothstein, Ira Z.},
    title = {Renormalization group scaling in nonrelativistic QCD},
    journal = {Phys. Rev. D},
    volume = {61},
    pages = {074025},
    year = {2000},
    doi = {10.1103/PhysRevD.61.074025},
    eprint = {hep-ph/9910209},
    archivePrefix = {arXiv},
    primaryClass = {hep-ph},
    reportNumber = {UCSD-PTH-99-11},
    SLACcitation = {%%CITATION = HEP-PH/9910209;%%}
}

@article{Manohar:2000a,
    author = {Manohar, Aneesh V. and Stewart, Iain W.},
    title = {Renormalization group analysis of the QCD quark potential to order $v^2$},
    journal = {Phys. Rev. D},
    volume = {62},
    pages = {014033},
    year = {2000},
    doi = {10.1103/PhysRevD.62.014033},
    eprint = {hep-ph/9912226},
    archivePrefix = {arXiv},
    primaryClass = {hep-ph},
    SLACcitation = {%%CITATION = HEP-PH/9912226;%%}
}

@article{Manohar:2000b,
    author = {Manohar, Aneesh V. and Stewart, Iain W.},
    title = {The QCD heavy quark potential to order $v^2$: One loop matching conditions},
    journal = {Phys. Rev. D},
    volume = {62},
    pages = {074015},
    year = {2000},
    doi = {10.1103/PhysRevD.62.074015},
    eprint = {hep-ph/0003032},
    archivePrefix = {arXiv},
    primaryClass = {hep-ph},
    SLACcitation = {%%CITATION = HEP-PH/0003032;%%}
}

@article{Hoang:2003a,
    author = {Hoang, Andre H. and Manohar, Aneesh V. and Stewart, Iain W. and Teubner, Thomas},
    title = {The $t\bar{t}$ threshold cross section at NNLL order},
    journal = {Phys. Rev. D},
    volume = {65},
    pages = {014014},
    year = {2002},
    doi = {10.1103/PhysRevD.65.014014},
    eprint = {hep-ph/0107144},
    archivePrefix = {arXiv},
    primaryClass = {hep-ph},
    SLACcitation = {%%CITATION = HEP-PH/0107144;%%}
}

@article{Hoang:2003b,
    author = {Hoang, Andre H. and Stewart, Iain W.},
    title = {Ultrasoft renormalization in nonrelativistic QCD},
    journal = {Phys. Rev. D},
    volume = {67},
    pages = {114020},
    year = {2003},
    doi = {10.1103/PhysRevD.67.114020},
    eprint = {hep-ph/0209340},
    archivePrefix = {arXiv},
    primaryClass = {hep-ph},
    reportNumber = {MPI-PhT-2002-49, INT-PUB-02-46},
    SLACcitation = {%%CITATION = HEP-PH/0209340;%%}
}

@article{Hoang:2004,
    author = {Hoang, Andre H.},
    title = {Three-loop anomalous dimension of the heavy quark pair production current in non-relativistic QCD},
    journal = {Phys. Rev. D},
    volume = {69},
    pages = {034009},
    year = {2004},
    doi = {10.1103/PhysRevD.69.034009},
    eprint = {hep-ph/0307376},
    archivePrefix = {arXiv},
    primaryClass = {hep-ph},
    reportNumber = {MPP-2003-38},
    SLACcitation = {%%CITATION = HEP-PH/0307376;%%}
}

@article{Rothstein:2018,
    author = {Rothstein, Ira Z. and Shrivastava, Prashant and Stewart, Iain W.},
    title = {Manifestly soft gauge invariant formulation of vNRQCD},
    journal = {Nucl. Phys. B},
    volume = {939},
    pages = {405--428},
    year = {2019},
    doi = {10.1016/j.nuclphysb.2018.12.027},
    eprint = {1806.07398},
    archivePrefix = {arXiv},
    primaryClass = {hep-ph},
    reportNumber = {MIT-CTP-4892},
    SLACcitation = {%%CITATION = ARXIV:1806.07398;%%}
}

@incollection{Hoang:2002,
    author = {Hoang, Andre H.},
    title = {Heavy quarkonium dynamics},
    booktitle = {At the Frontier of Particle Physics: Handbook of QCD},
    editor = {Shifman, M.},
    volume = {4},
    pages = {2215--2331},
    publisher = {World Scientific},
    address = {Singapore},
    year = {2002},
    doi = {10.1142/9789812777270_0002},
    eprint = {hep-ph/0204299},
    archivePrefix = {arXiv},
    primaryClass = {hep-ph},
    reportNumber = {MPI-PhT-2002-14},
    SLACcitation = {%%CITATION = HEP-PH/0204299;%%}
}

@article{Peset:2015vvi,
    author = "Peset, Clara and Pineda, Antonio and Stahlhofen, Maximilian",
    title = "{Potential NRQCD for unequal masses and the B$_{c}$ spectrum at N$^{3}$LO}",
    eprint = "1511.08210",
    archivePrefix = "arXiv",
    primaryClass = "hep-ph",
    reportNumber = "DESY-15-223, MITP-15-107",
    doi = "10.1007/JHEP05(2016)017",
    journal = "JHEP",
    volume = "05",
    pages = "017",
    year = "2016"
}

@phdthesis{Pan:2023thesis,
    author = {Pan, Jichen},
    title = {Application of Non-relativistic Quantum Chromodynamics and 1+1 Dimensional HQET},
    school = {University of Chinese Academy of Sciences},
    type = {Ph.D. thesis},
    address = {Beijing, China},
    month = jun,
    year = {2023},
    note = {Institute of High Energy Physics, Chinese Academy of Sciences}
}
\end{document}